\documentclass[prb,twocolumn,superscriptaddress]{revtex4}

\usepackage{amsmath,amssymb,amsfonts,amsthm}    

\usepackage{braket}
\usepackage{enumerate}
\usepackage[pdftex]{graphicx} 
\usepackage{braket}
\usepackage{epstopdf}
\usepackage{amsmath, amssymb, amsfonts, bm, textcomp}
\usepackage{float, placeins, tikz, graphicx}
\usepackage{multirow, array}

\newcommand{\te}[1]{{\text{#1}}}

\begin{document}

\bibliographystyle{apsrev4-1} 

\title{Tuning the critical solution temperature of polymers by copolymerization}

\author{Bernhard Schulz}
\affiliation{Institut f\"ur Weiche Materie und Funktionale Materialien, Helmholtz-Zentrum Berlin, 14109 Berlin, Germany}
\affiliation{Institut f\"ur Physik, Humboldt-Universit{\"a}t zu Berlin, 12489 Berlin, Germany}
\author{Richard Chudoba}
\affiliation{Institut f\"ur Weiche Materie und Funktionale Materialien, Helmholtz-Zentrum Berlin, 14109 Berlin, Germany}
\affiliation{Institut f\"ur Physik, Humboldt-Universit{\"a}t zu Berlin, 12489 Berlin, Germany}
\author{Jan Heyda}
\affiliation{Department of Physical Chemistry, University of Chemistry and Technology, Prague, 166 28 Praha 6, Czech Republic}
\author{Joachim Dzubiella}
\email{joachim.dzubiella@helmholtz-berlin.de}
\affiliation{Institut f\"ur Weiche Materie und Funktionale Materialien, Helmholtz-Zentrum Berlin, 14109 Berlin, Germany}
\affiliation{Institut f\"ur Physik, Humboldt-Universit{\"a}t zu Berlin, 12489 Berlin, Germany}

\begin{abstract}

We study statistical copolymerization effects on the upper critical solution temperature (CST) of generic homopolymers by means of coarse-grained Langevin dynamics computer simulations and mean-field theory. Our systematic investigation reveals that the CST can change monotonically or non-monotonically with copolymerization, as observed in experimental studies, depending on the degree of non-additivity of the monomer (A-B) cross-interactions. The simulation findings are confirmed and qualitatively explained by a combination of a two-component Flory-de Gennes model for polymer collapse and a simple thermodynamic expansion approach. Our findings provide some rationale behind the effects of copolymerization and may be helpful for tuning CST behavior of polymers in soft material design.

\end{abstract}

\maketitle

\section{Introduction}

Thermoresponsive polymers can strongly respond to chemical and physical stimuli in their environment and have been therefore heavily investigated recently  for their use as 'smart' and adaptable functional materials.~\cite{stuart}  If the stimulus is strong enough or the system is close to its critical solution  temperature (CST), a volume phase transition of the polymer is triggered that significantly alters all physicochemical properties  of the material.  For this reason, the application of thermosensitive polymers is explored, for example,  in the field of sensors, \cite{uchiyama2004fluorescent,hu1998polymer,liu2014thermoresponsive} separation and filtration systems,~\cite{kanazawa1996temperature,kikuchi2002intelligent, kanazawa2007thermally,tan2012thermoresponsive} as well as drug carriers and tissue engineering.~\cite{liu2010recent,ward2011thermoresponsive,schmaljohann2006thermo} 
 It is therefore of high interest to be able to tailor the CST according to the needs of the desired application.
 
It is well known that the CST can be tuned by statistical copolymerization, i.e., by replacing monomers of
a homopolymer by a second monomer type in a random, or statistically repeated fashion.  The CST has been observed then to depend either monotonically~\cite{feil1993effect, liu1999, djokpe2001, keerl, maeda2009, seuring2010, park2007, plamper2012, glassner2014} or non-monotonically \cite{park2007, keerl, plamper2012,maeda2009} on the degree of copolymerization. In the typical scenario of a lower CST (LCST) in aqueous solvents, monotonic trends are often qualitatively explained by  the degree of hydrophobicity/philicity of the copolymerizing monomers.~\cite{liu1999}  Copolymerizing with hydrophobic monomers leads to a lower LCST~\cite{seuring2010} because the overall hydrophobic attraction increases and leads to a collapse  already at lower temperature, while incorporating hydrophilic monomers analogously  increases the LCST.~\cite{djokpe2001,feil1993effect} However, this explanation fails  to describe non-monotonic trends in the CST behavior.  
 
The typical theoretical approach to describe critical solution temperatures of polymer systems essentially revolves around extensions of 
Flory-Huggins (FH) lattice theory,~\cite{flory1953principles, deGennes1979} for instance, as frequently applied to copolymer blends.~\cite{bopp, brinke, barlow, freed, dawson} Noteworthy is here the work of Paul and Barlow~\cite{barlow} who demonstrated that the existence of non-monotonic copolymer miscibility ('miscibility windows') sensitively depends on the degree of additivity of the monomeric interaction energies between the components. 
In another line of work directly applied to the LCST of a thermosensitive copolymer in water, Kojima and Tanaka~\cite{tanaka2013} 
theoretically explained the nonlinear depression of the LCST of thermosensitive polymers by a combination of FH theory and  
a microscopic level approach based  on cooperative hydration effects. 

In a related but different view, the CST for a one-component polymer system in solvent can also be regarded as being signified by a  collapse transition (or coil-to-globule) of a long, single polymer.~\cite{brochard, baysal} This perspective is taken in this work.  In contrast to the FH lattice theory, single polymer collapse is  typically described within a mean-field Flory-de Gennes approach.~\cite{flory1953principles, deGennes1979, rubinstein}  In the classical homopolymer case, the free energy $F(R)$ of a chain with $N$ monomers is the sum of ideal chain entropy $\sim R^2/Nb^2$  and the mean-field monomer interactions expressed by a virial expansion in monomer density $\sim N/R^3$, with $R$ describing the mean polymer size and $b$ the segment length. 
On the simplest level, the free energy reads~\cite{deGennes1979, rubinstein}
\begin{eqnarray}
  F(R) \sim R^2/Nb^2 + B_2N^2/R^3 + B_3N^3/R^6, \label{eq:OneComponentFdG}
\end{eqnarray}
where $B_2$ and $B_3$ are the second and third virial coefficients of the monomer gas, respectively. However, this one-component 
approach seems not be adequate to describe the effects of copolymerization on polymer size as effective virial coefficients are 
used instead of those distinguishing between the monomer self- and cross-interactions.  Supposedly, as in the 
FH lattice models for polymer blends, therefore both polymer components need to be considered with individual effective pair interactions. 
In contrast to FH approach to polymer blends, these interactions are 'effective' for solvated copolymers in the sense that the 
solvent degrees of freedom are integrated out and their effects are included in the effective potentials. 

In the present work, we explore copolymerization effects on the collapse transition of a single polymer  by coarse-grained, implicit-solvent Langevin dynamics 
computer simulations and a two-component Flory-de Gennes Model. For this, we use the most generic polymer model in which the polymer is modeled as a  freely jointed chain where individual monomers are interacting via a Lennard-Jones potential. The degree of copolymerization of a homopolymer with monomer of type A is then expressed in percentage of statistically placed monomers of type B.  Since we employ temperature-independent effective pair interactions between the monomers, we thus focus on the behavior of the upper CST (UCST).  
We first demonstrate that both simulation and Flory theory can reproduce all experimental trends. A thermodynamic expansion approach of the transition (Flory) free energy  is employed to  interpret the changes of the UCST upon copolymerization based on the dominating monomer pair interactions.  
The non-monotonic behavior of the UCST can then be traced back to non-additive cross-interactions
between the monomers, resembling the behavior of miscibility windows in polymer blends.~\cite{barlow}  
This picture is consistent with experimental work on the LCST of thermosensitive polymers, where preferential hydrogen bonding between the unlike 
NIPAM and N,N-diethylacrylamide (DEAAM) monomers is reported, and it was pointed out that this is an intramolecular phenomenon, 
hardly observed for homopolymeric mixtures.~\cite{plamper2012} 

We note that our approaches and interpretation should be equally valid, however, for LCST behavior, at least on a qualitative level.  
In this case,  temperature-dependent effective pair interactions have to be introduced in the simulations. Hydrophobic interactions 
lead to more attractive pair potentials with increasing temperature,~\cite{schellman, paschek, makowski} reflected by $B_2$-values that 
decrease with rising temperature.~\cite{reinhardt2013fine} These effects will drive collapse in both simulation and Flory 
theory for increasing temperature as characteristic for LCST behavior.  If such a pair potential picture is sufficient for a quantitative 
treatment of the thermodynamics of the collapse transition at the LCST, however, is yet unclear and 
needs further consideration in future work. 

\section{Coarse-grained Langevin computer simulations\label{sec:Simulations}}

\subsection{Simulation model and setup\label{subsec:Setup}}

Our simulations are based on a generic freely jointed chain model,~\cite{compsim} where copolymers composed of two different monomer species, A and B,  are investigated. The total number of monomers is constant and given by $N = N_A + N_B = 100$, where $N_A$ and $N_B$ refer to the number of monomers of species A and B, respectively. The degree of copolymerization $\chi = N_B/N$ 
is given in percentage of monomer B; hence $\chi=0 $ corresponds to a homopolymer with $0 \%$ of monomer B (100\% of monomer A), and $\chi=1$ corresponds  to a homopolymer with $100 \%$ of monomer B. Our statistical copolymerization is always performed in a periodically repeated fashion, that is, all B monomers are placed in equidistant places along the polymer. A value $\chi = 0.1$, for instance, thus corresponds to a (AAAAAAAAAB)$_{10}$ repeat,  
while $\chi = 0.5$  corresponds to the alternating (AB)$_{50}$ repeat.  All monomers interact through the 
Lennard-Jones (LJ) potential
\begin{eqnarray}
  V_{ij}(r)=4\epsilon_{ij}[(\sigma_{ij}/r)^{12} - (\sigma_{ij}/r)^6], \label{eq:LJ} 
\end{eqnarray}
where $\epsilon_{ij}$ and $\sigma_{ij}$ set the energy scale and length scale, respectively, and $i,j = A,B$. 
The LJ parameters used for the investigated copolymers are summarized in Tables~\ref{tab:setup} and \ref{tab:setup_non_Lorentz}. 
In the former table, the cross interaction obeys the Lorentz-Berthelot mixing rule, which states that $\epsilon_{AB}=\sqrt{\epsilon_{AA}\epsilon_{BB}}$ and $\sigma_{AB}=(\sigma_{AA}+\sigma_{BB})/2$ holds, that is, the interactions are additive. In the second set of polymer systems, the additivity is violated and
we employ cross energies  $\epsilon_{AB}\neq\sqrt{\epsilon_{AA}\epsilon_{BB}}$ that are either smaller or larger than the geometrical mean, while
holding the interaction lengths $\sigma_{AA} = \sigma_{BB} = \sigma_{AB}$ fixed. Chemically, the loosening of the additivity restriction could correspond to A and B monomers that interact similarly among themselves, e.g., repulsive or only weakly attractive, but more strongly attract the other type. An example could be donor- and acceptor molecules that preferentially form hydrogen bonds between each other but effectively repel the same of its kind, as realized for N-isopropylacrylamide (NIPAM) and N,N-diethylacrylamide (DEAAM).~\cite{plamper2012}

\begin{table}[b]
\caption{Summary of the investigated {\it additive} monomeric interactions. Monomer A and B interact via a Lennard-Jones potential employing the  Lorentz-Berthelot mixing rules $\epsilon_{AB}=\sqrt{\epsilon_{AA}\epsilon_{BB}}$ and $\sigma_{AB}=(\sigma_{AA}+\sigma_{BB})/2$. The unit of $\epsilon_{ij}$ is given in [kJ/mol] and the unit of $\sigma_{ij}$ in [nm].\label{tab:setup}}
\begin{ruledtabular}
\begin{tabular}{c c c c}
 $\epsilon_{AA}$ & $\sigma_{AA}$  &  $\epsilon_{BB}$ & $\sigma_{BB}$ \\ 
 \hline
 $1.00$                    & $0.30$                 &  $0.50$                  & $0.30$    \\            
 $1.00$                    & $0.30$                 &  $1.50$                  & $0.30$    \\                            
 $1.00$                    & $0.30$                 &  $0.50$                  & $0.45$    \\                            
 $1.00$                    & $0.30$                 &  $1.00$                  & $0.45$    \\                            
 $1.00$                    & $0.30$                 &  $1.50$                  & $0.45$    \\                            
\end{tabular}
\end{ruledtabular}
\vspace*{0.5cm}
\caption{Summary of the investigated {\it non-additive} monomeric interactions. Here, the interaction cross parameter $\epsilon_{AB}$ is {\it not} given via the Lorentz-Berthelot mixing rule. A-A and B-B interactions are the same. The unit of $\epsilon_{ij}$ is given in [kJ/mol] and for $\sigma$ in [nm].\label{tab:setup_non_Lorentz}}
\begin{ruledtabular}
\begin{tabular}{c c c c }
 $\epsilon_{AA}\equiv\epsilon_{BB}$  & $\sigma_{AA}\equiv\sigma_{BB}$  &  $\epsilon_{AB}$  & $\sigma_{AB}$  \\ 
 \hline
$1.50$                    & $0.30$                 &  $1.00$                  & $0.30$    \\            
$1.50$                    & $0.45$                 &  $1.00$                  & $0.45$    \\            
$1.50$                    & $0.30$                 &  $1.25$                  & $0.30$    \\            
$1.50$                    & $0.45$                 &  $1.25$                  & $0.45$    \\            
$1.50$                    & $0.30$                 &  $1.75$                  & $0.30$    \\            
$1.50$                    & $0.45$                 &  $1.75$                  & $0.45$    \\            
$1.50$                    & $0.30$                 &  $2.00$                  & $0.30$    \\            
$1.50$                    & $0.45$                 &  $2.00$                  & $0.45$    \\            
\end{tabular}
\end{ruledtabular}
\end{table}

The stochastic Langevin simulations are performed in a NVT ensemble in a cubic simulation box with side lengths of $L_x=L_y=L_z=30$ nm. The GROMACS simulation package \cite{gromacs2005} is used to integrate Langevin's  equation of motion. A time step of 10~fs is used. The center of mass translation and rotation is removed every tenth step. The bond length between neighboring monomers is set to $b=0.38$~nm, and is constrained by the LINCS algorithm as implemented in GROMACS. A friction constant of $\xi = 0.5\, \te{ps}^{-1}$ is employed. Moreover, the LJ-interaction between two neighboring monomers within the chain is excluded, and the LJ-interactions between all other monomers  is calculated within a cut-off distance of $1.2$~nm. All simulations are performed with an implicit solvent. The degree of copolymerization of initially $\chi=0 $ (A homopolymer)  is explored up to $\chi=1$ (B homopolymer) in 0.1 increments. For every single $\chi$, a temperature range from 75 to 675~K (in steps of $25$~K) is investigated. Every system has been simulated for 500~ns after 50~ns of equilibration.

\subsection{Definition of the critical solution temperature\label{sec:T_c}}

\begin{figure}[htb]
    \includegraphics[width=9cm]{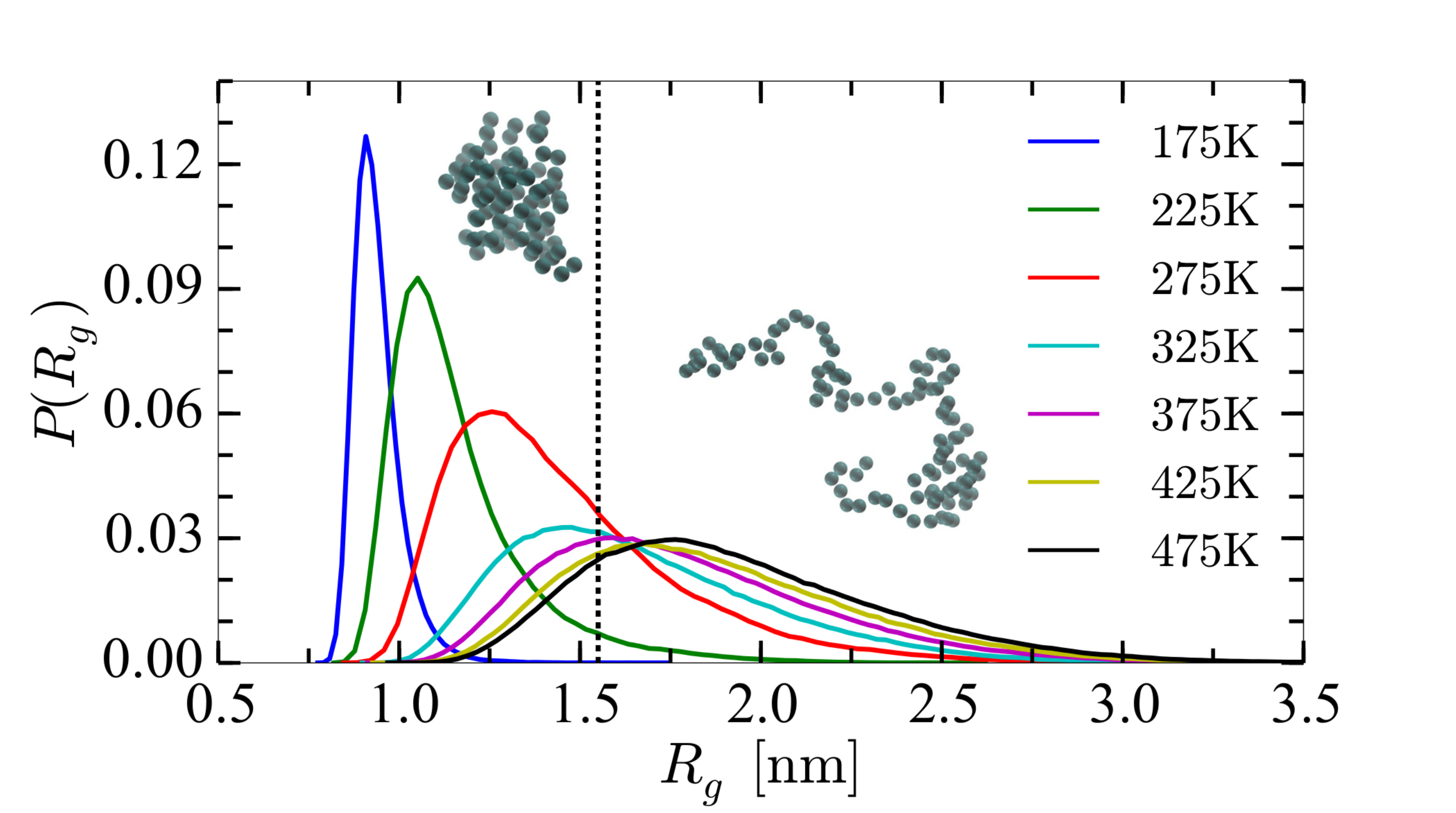}
    \caption{Probability distribution $P(R_g)$ of the polymer size in terms of the radius of gyration $R_g$ for different temperatures $T$ (see legend) calculated in the coarse-grained simulations. The degree of copolymerization is $\chi=0$, and the LJ interaction parameters are $\sigma_{AA}=0.3$~nm, $\epsilon_{AA}=1$~kJ/mol. The threshold size that separates collapsed and swollen states (see the inset illustrations) is $R_g^\te{crit}=1.55$~nm, depicted by the vertical dashed line. \label{fig:ProbDist}}
\end{figure}

It is important to emphasize that a first-order CST is only defined for polymer solutions in the thermodynamic limit~\cite{ivanov1998finite, schnabel2011} for polymers with internal degrees of freedom,~\cite{maffi} and therefore it is difficult to obtain the CST from our finite-size simulation using a minimalistic polymer model. 
However, we can define and calculate a critical transition temperature, $T_c(\chi)$, 
which is strongly related but not equal to the CST, to discuss the qualitative trends with changing copolymerization $\chi$.
For this please consider Fig.~\ref{fig:ProbDist}, where the population probability distribution $P(R_g)$ of the radius of gyration $R_g$ is shown for a homopolymer ($\chi=0$) at different temperatures $T$. Obviously, $R_g$ is a function of temperature $T$ (as well as a function of the  degree of copolymerization $\chi$ as we show later), thus $P=P(R_g;T,\chi)$. 
It can be clearly seen that the polymer favors the extended (coil like) state for high temperatures, which is a clear sign for an upper CST (UCST)-like behavior. This behavior is found to be a universal feature of all polymers defined in Table~\ref{tab:setup} and Table~\ref{tab:setup_non_Lorentz} as we employ temperature-independent pair potentials. As discussed in literature, e.g., on thermosensitive polymer systems,~\cite{reinhardt2013fine} 
an explicitly temperature-dependent effective pair potential is  needed (typically originating
from hydrophobic interactions~\cite{schellman, paschek, reinhardt2013fine, makowski}) to obtain a lower CST (LCST) behavior.

We can now define the critical temperature $T_c(\chi)$ as the temperature of the state at which the probability to find the copolymer in a collapsed globule (g) state $P_g$ is equal to the probability to find the copolymer in a swollen coil (c) state $P_c$, hence $P_g=P_c$ at $T=T_c$. We define the population probabilities by averages of  $P(R_g)$, via 
\begin{align}
  P_g = \int_{R_g\le R_g^\te{crit}} P(R_g)dR_g \nonumber  \\
  P_c =  \int_{R_g > R_g^\te{crit}} P(R_g)dR_g, \label{eq:Prob_Integral}
\end{align}
where $R_g^\te{crit}$ is an arbitrarily (but motivated) chosen threshold value. A reasonable choice for $R_g^\te{crit}$ is the radius of gyration of an ideal polymer (i.e., describing the chain size in a $\theta$-solvent) which is given by $\sqrt{N}b/\sqrt{6}$. Hence, to explore transitions in our simulations, we set  $R_g^\te{crit}=R_g^\te{ideal}=\sqrt{N}b/\sqrt{6}=1.55$~nm. We note that the qualitative trends of $T_c(\chi)$ are quite insensitive to the exact value of
$R_g^\te{crit}$ and even remain valid if the distribution of the end-to-end extension of the polymer is evaluated as a measure of polymer
size.

\section{Mean-field and thermodynamic descriptions of the CST\label{sec:Polymer_Models}}

In this section a theoretical mean-field approach for calculating the critical temperature $T_c(\chi)$ is presented based on a two-component 
Flory-de Gennes model.~\cite{flory1953principles, deGennes1979, heyda2013}  Subsequently, we discuss a recently introduced thermodynamic  expansion 
approach of the transition free energy~\cite{heyda2014a,heyda2014} that relates copolymerization-induced  free energy changes to 
small changes of the CST and serves for some interpretation of the simulation data. 

\subsection{Two-component Flory-de Gennes Model\label{subsec:Flory_de_Gennes}}

In the mean-field Flory-de Gennes picture the monomer-monomer interactions are described by second and third-order virial coefficients. In contrast to the one-component approach, cf.~eq \eqref{eq:OneComponentFdG},  the copolymerization degrees of freedom are explicitly considered as the second component, which implies that we define three independent second virial coefficients $B_2^{AA}$, $B_2^{BB}$, and $B_2^{AB}$. These virial coefficients describe the monomer self-interactions (AA and BB)  and the monomer cross-interactions (AB), respectively.  These are in principle effective interactions as solvent degrees of freedom are integrated out. In our comparison to implicit-solvent computer simulations, these are given by the LJ pair potential. The resulting expression for the mean-field (Helmholtz) free energy  is strongly related to the one recently introduced describing cosolute-induced polymer swelling and collapse,~\cite{heyda2013} and reads
\begin{align}
\beta  F_{\rm mf}(R) & = \frac{3R^2}{2Nb^2} + \frac{\pi^2 N b^2}{12 R^2} +  V \sum_{i,j=A,B} \rho_i \rho_j B_2^{ij} \nonumber \\
       & +  \frac{V}{2} \sum_{i,j,k=A,B} \rho_i \rho_j \rho_k B_3^{ijk}, \label{eq:TwoComponentFdG}
\end{align}
where $\beta = 1/(k_B T)$ is the inverse thermal energy and we used the index 'mf'  as subscript to the free energy to signify the 
mean-field character of this treatment.  The second term represents confinement entropy of the polymer in the collapse state.~\cite{rubinstein}
As usual in Flory theory, the end-to-end distance $R$ serves to represent the mean size of the polymer (irrespective if it is coil or globule) in an equilibrium state, and with $V =4\pi R^3/3 $ we approximate the spherical volume occupied by the polymer that has to be considered configurationally averaged in the mean-field approach. The number density $\rho_i = N_i/V$ is  then mean monomer density of monomer species $i$. The densities are directly related to the copolymerization parameter by $\rho_A = \rho_N(1-\chi)$ and $\rho_B = \rho_N\chi$, where $\rho_N = N/V$ is the total monomer density in the sphere occupied on average by the polymer.  $B_2^{ij}$ is the second and $B_3^{ijk}$ the third virial coefficient for the respective monomer combinations. Due to the mean-field nature of this  treatment it holds for both 
statistically repeated and random copolymerization as long as the polymer is long enough, that is, $N$ is much larger than a typical repeat unit 
or the correlation number between copolymerizing monomers. 

\begin{figure}[bht]
          \includegraphics[width=8cm]{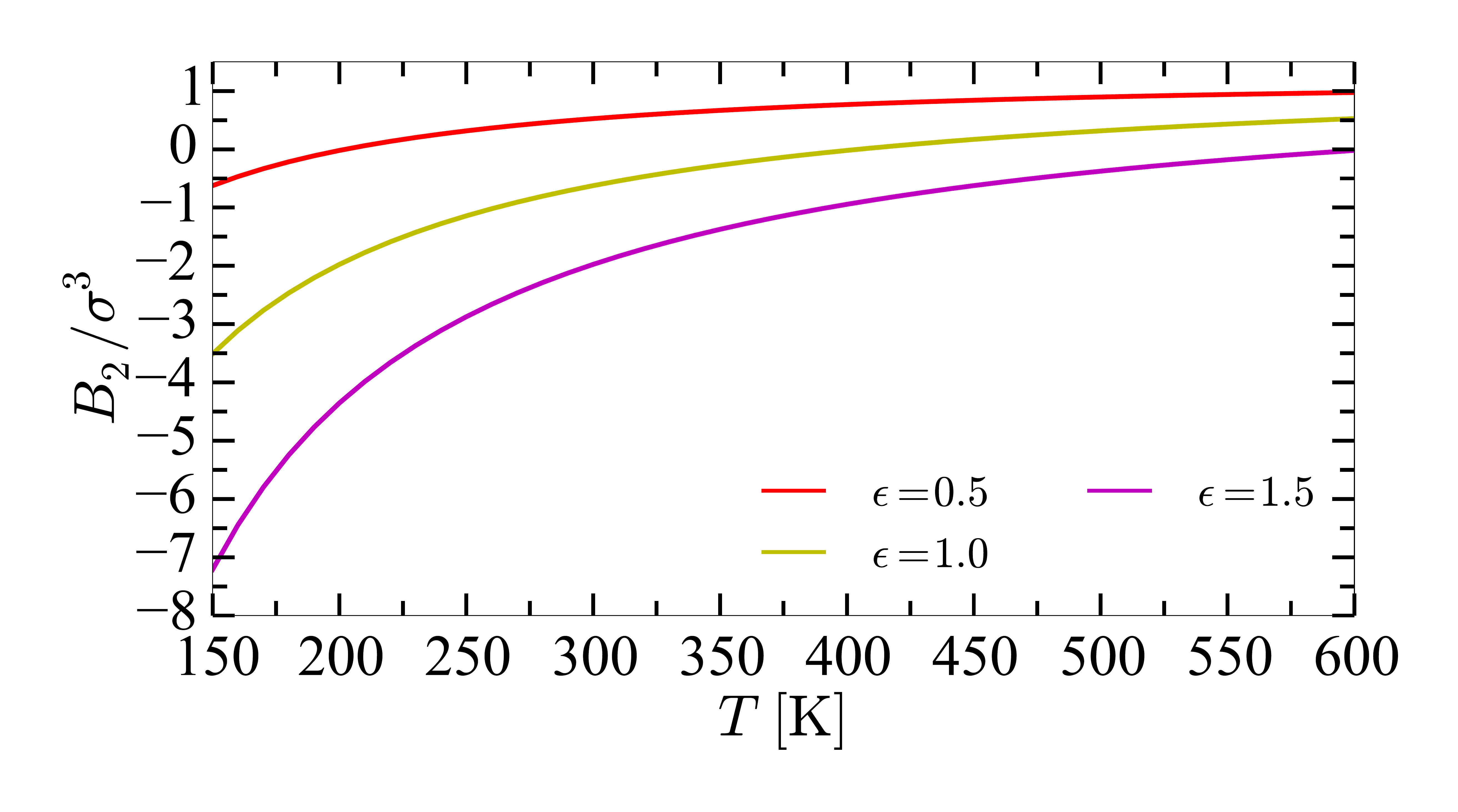}
          \includegraphics[width=8cm]{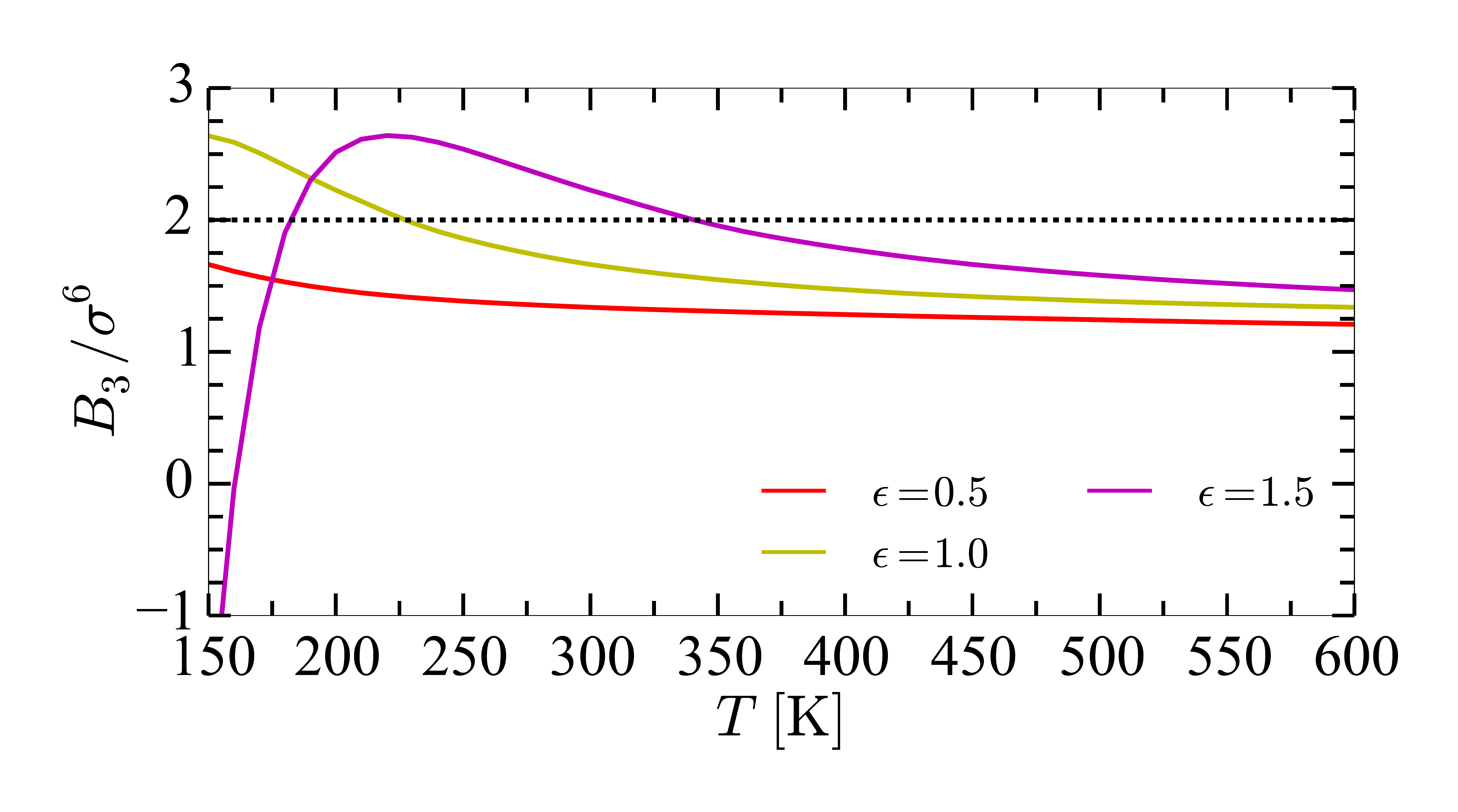}
    \caption{(top) Second virial coefficient $B_2$ of the LJ interaction as a function of temperature $T$ with LJ parameter $\epsilon$ as shown in the legend. (bottom) Third virial coefficient $B_3$ of the LJ interactions as a function of $T$ with LJ parameters shown in the legend. The dotted line depicts the hard sphere case approximation with $B_3^{ijk}=2\sigma^6$.  
    Units of $\epsilon$ are in [kJ/mol], and $\sigma$ is the LJ size parameter.  
    \label{fig:Virial_coefficients}}
\end{figure}

The second virial coefficient is explicitly defined via 
\begin{align}
  B_2^{ij}(T)=-\frac{1}{2}\int d^3r \left[ \exp(-\beta V_{ij}(r)) - 1 \right], \label{eq:SecondVirial}
\end{align}
where $V_{ij}(r)$ represents the LJ potential.  For $B_2>0$ the pair interaction is said to be repulsive and for $B_2<0$ attractive. Due to the strong excluded-volume repulsion for small monomeric separations, there is a turnover form 
attractive to repulsive for high enough temperatures, as shown in Fig.~\ref{fig:Virial_coefficients} (top), where we plot the $B_2$ coeffcient for the LJ  
potential versus $T$ for various values of the LJ $\epsilon$ parameter. The third virial coefficient $B_3^{ijk}$, also calculated explicitly as shown in Fig.~\ref{fig:Virial_coefficients} (bottom),  is positive for nearly all temperatures. It actually turns out that for our purposes the action of 3-body effects is well approximated by the virial coefficient 
for hard spheres for all temperatures. It is given by $B_3^{ijk}=2\sigma^6$ (see the dotted line in Fig.~2 bottom), where the LJ-size $\sigma$ is the approximate excluded-volume radius for
the corresponding monomer pair potential. 

The equilibrium radius $R_{eq}$ in the Flory theory is obtained by minimizing eq.~\eqref{eq:TwoComponentFdG} with respect to the polymer size $R$ for a fixed $T$ and $\chi$.  The CST in our definition is calculated by finding the temperature at which the probabilities of being in the globular and coil states, 
\begin{eqnarray}
P_g=\int_{R\le R_\te{crit}} e^{-\beta F_{\rm mf}(R)} dR \nonumber \\
P_c=\int_{R\ge R_\te{crit}} e^{-\beta F_{\rm mf}(R)} dR
\end{eqnarray}
respectively, are equal. This definition is analogous to eq.~(3) using $P(R) = \exp(-\beta F_{\rm mf}(R))$. 

For small perturbations by copolymerization, we can utilize eq.~\eqref{eq:TwoComponentFdG} to derive a simple equation for the change of
the transition temperature, $\Delta T_c$,  in the next section that allows an interpretation  of the dominating factors that change $T_c$ with $\chi$.  
As a prerequisite, we Taylor-expand  eq.~\eqref{eq:TwoComponentFdG} in $\chi$ around the reference 
configuration at $\chi=0$.  In first order in $\chi$, we obtain per monomer
\begin{align}
    \frac{1}{N}\frac{\partial F_{\rm mf}}{\partial \chi}\biggm\rvert_{\chi=0} = 2k_BT\rho_N\left( B_2^{AB} - B_2^{AA} \right) \nonumber \\
    + \frac{3}{2}k_BT\rho_N^2\left(B_3^{AAB}-B_3^{AAA} \right),
    \label{Flory_linear}
\end{align}
that is, changes in the free energy for small perturbations originate from differences in the virial coefficients between the homopolymeric A-state
and the first order perturbed state. Analysis of the simulations data actually  shows that the $B_2$-contributions in the collapsed state 
are dominating in most of our examples, as the density of the collapsed globule (g) state, $\rho_g \gg \rho_c$, is significantly higher than the one of the swollen coil (c) state, and $B_3$ contributions are small or cancel each other for small perturbations.   
The major changes in the free energy in eq.~(\ref{Flory_linear}) from perturbing by a small copolymerization $\chi\ll1$
are thus essentially given by pair energies in the collapsed state, expressed by 
\begin{align}
 \Delta F_{\rm mf} \approx 2k_BTN\rho_g\left( B_2^{AB} - B_2^{AA} \right)\chi.
    \label{Flory_linear2}
\end{align}
Hence, the crucial quantity which determines the slope of the linear free energy change is the difference in interactions provided by 
 $B_2^{AB}-B_2^{AA}$. This is similar to the effective Flory interaction energy parameter in Flory-Huggins lattice approaches to 
 polymer mixtures.~\cite{flory1953principles, deGennes1979, bopp, brinke, barlow, freed, dawson} However, it is more general 
 as it includes also the effects of varying excluded-volume interactions and van der Waals attraction, which is not so easy to consider in lattice models. 
 In the following, we combine this linear analysis with an insightful thermodynamic definition of the change in CST for small 
temperature changes ($\Delta T_c \ll T_c$)  that serves for further interpretation of the simulation data by the 
Flory approach.

\subsection{Thermodynamic Expansion of the Two-State Free Energy\label{subsection:Thermodynamic_expansion_section}}

The following approach is based on  a previously introduced thermodynamic expansion model to describe charge or cosolute effects on the CST.~\cite{heyda2014a,heyda2014}   Here, it is assumed that the copolymer transition at the CST can be understood as a transition from a dense collapsed 
globule  to an expanded, coil-like state in a bimodal free energy landscape~\cite{baysal,ivanov1998finite, maffi} 
as a function of the copolymer specific volume.   At the transition state,  $T_c$ is defined as the state where $P_g = P_c$ holds. 
The transition free energy between the two states is then
\begin{eqnarray}                                                                                             
\Delta F(T_c,\chi)&=&F_\te{g}-F_\te{c}=-k_BT\ln{(P_g/P_c)} 
\end{eqnarray}
 and thus vanishes at $T_c(\chi)$, that is, $\Delta F(T_c)=0$. In order to relate how a change of the transition free energy, defined as 
 $\Delta\Delta F = \Delta F(T_c, \chi>0) - \Delta F(T_c,0)$, relates to the change of the CST, $\Delta T_c(\chi)$, we perform 
 a Taylor-expansion of the transition free energy $\Delta F$ with respect to the temperature $T$ and copolymerization $\chi$ 
 around a  thermodynamic reference state up to first order. 
 A reasonable reference state is $T=T_c$ at $\chi=0$. The expansion then reads
\begin{eqnarray}
    \Delta F(T_c+\Delta T_c,\chi) &\approx& \Delta F(T_c,0)+\left( \frac{\partial \Delta F(T,\chi)}{\partial T}\right)_{T_c,0}\Delta T_c 
    \nonumber \\ 
    &+& \left( \frac{\partial \Delta F(T,\chi)}{\partial \chi}\right)_{T_c,0}\chi + ...  
    \label{eq:Thermodynamic_expansion}
\end{eqnarray}
Here, $\Delta F(T_c,0) =0$ for the reference state.  Furthermore, 
we identify  the transition entropy $\Delta S_0=-({\partial \Delta F}/{\partial T})_{T_c,0}$ at the reference state. 
Note that for our relatively simple (implcit solvent) system $\Delta S_0 = S_c - S_s < 0$ is negative as the swollen state has a higher configurational entropy
than the collapsed state. 
The equation is then solved for $\Delta T_c$ with the condition that 
the change in critical temperature must satisfy $\Delta F(T_c+\Delta T_c,\chi)=0$. Considering this, 
identifying $(\partial \Delta F/\partial \chi)_{T_c,0}\chi = \Delta\Delta F(\chi)$,  and  
applying the new notation, we obtain
\begin{align}
    \Delta T_c(\chi) = \frac{\Delta\Delta F(\chi) }{\Delta S_0 } \simeq  \frac{2k_BTN\rho_g\left( B_2^{AB} - B_2^{AA} \right)\chi}{\Delta S_0}, 
    \label{eq:DeltaT_c}
\end{align}
for small perturbations $\chi\ll1$.  Eq.~\eqref{eq:DeltaT_c} therefore provides explicitly the change of the critical temperature 
$\Delta T_c(\chi)$ upon small copolymerization $\chi$. 
 Within the linear analysis of the Flory theory above, we see that the perturbation $\Delta \Delta F_{\rm mf}(T,\chi)$ is essentially given by the 
 pair interaction differences between A-A and A-B monomers in the globular state via eq.~(\ref{Flory_linear2}). The linear temperature change $\Delta T_c(\chi)$ is described by 
 this free energy change divided by the transition entropy at the reference state. For larger perturbations, non-linear
 behavior of   $\Delta T_c(\chi)$ can be generally expected as both $\Delta \Delta F(T,\chi)$ and the transition 
 entropy will depend nonlinearly on $\chi$. 

\section{Results}

\subsection{$T$ and $\chi$-dependence of polymer size \label{subsec:R_g}}

\begin{figure}[thb!]
        \includegraphics[width=8.5cm]{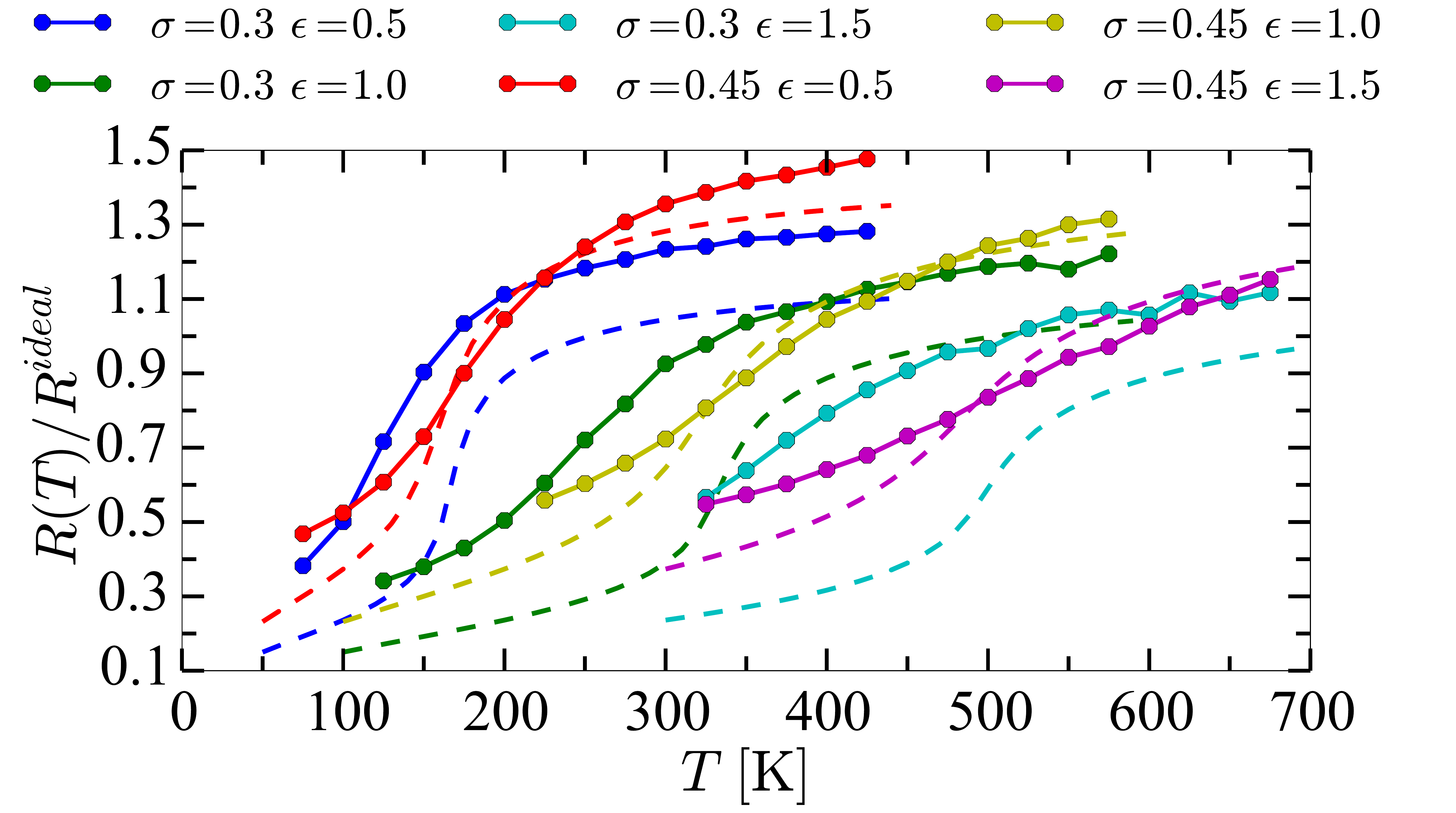}
    \caption{Scaled end-to-end distance $R/R^\te{ideal}$ as a function of temperature $T$ for various homopolymers ($\chi=0$) and LJ parameters as in the legend. Dashed lines are obtained from minimizing the Flory free energy eq.~(\ref{eq:TwoComponentFdG}), while symbols guided by solid lines depict simulation results. 
    \label{fig:R_T}}
\end{figure}

Let us first discuss the swelling and 'transition' behavior of our investigated homopolymers ($\chi=0$) in dependence of the temperature $T$. 
In the Flory theory the equilibrium size in terms of the end-to-end distance $R$ is obtained by minimizing the free energy $F(R;T)$ for a fixed $T$ (cf. Section \ref{subsec:Flory_de_Gennes}). We compare those results to the end-to-end distance calculated in the simulations in Fig.~\ref{fig:R_T}, where we 
plot them scaled by their respective ideal value $R^\te{ideal}$ as a function of $T$. All results show clearly that the polymers favor 
the extended (coil-like) state for high $T$, as expected for an UCST behavior. 

Both, simulation and theory results yield intermediate steep slopes of the resulting curves that indicate 
some higher order, continuous transition between collapsed and swollen states. For instance, the parameter set $\sigma = 0.3$~nm and $\epsilon=1.0$~kJ/mol shows a transition-like behavior between 300 and 400~K in the Flory theory (green dashed line).  We also see that the
transition region generally differs between simulation and Flory theory, pointing to some quantitative shortcomings
of the Flory theory. However, the overall simulation behavior is surprisingly very well captured, considering the simplicity and mean-field
character of the Flory approach, and we will focus on qualitative trends only.  We note that analogous curves for the $T$-dependence of the 
radius of gyration $R_g(T)$ extracted from simulations  show the same qualitative behavior, in particular the same transition region (not shown). 

\begin{figure}[h]
        \includegraphics[width=8cm]{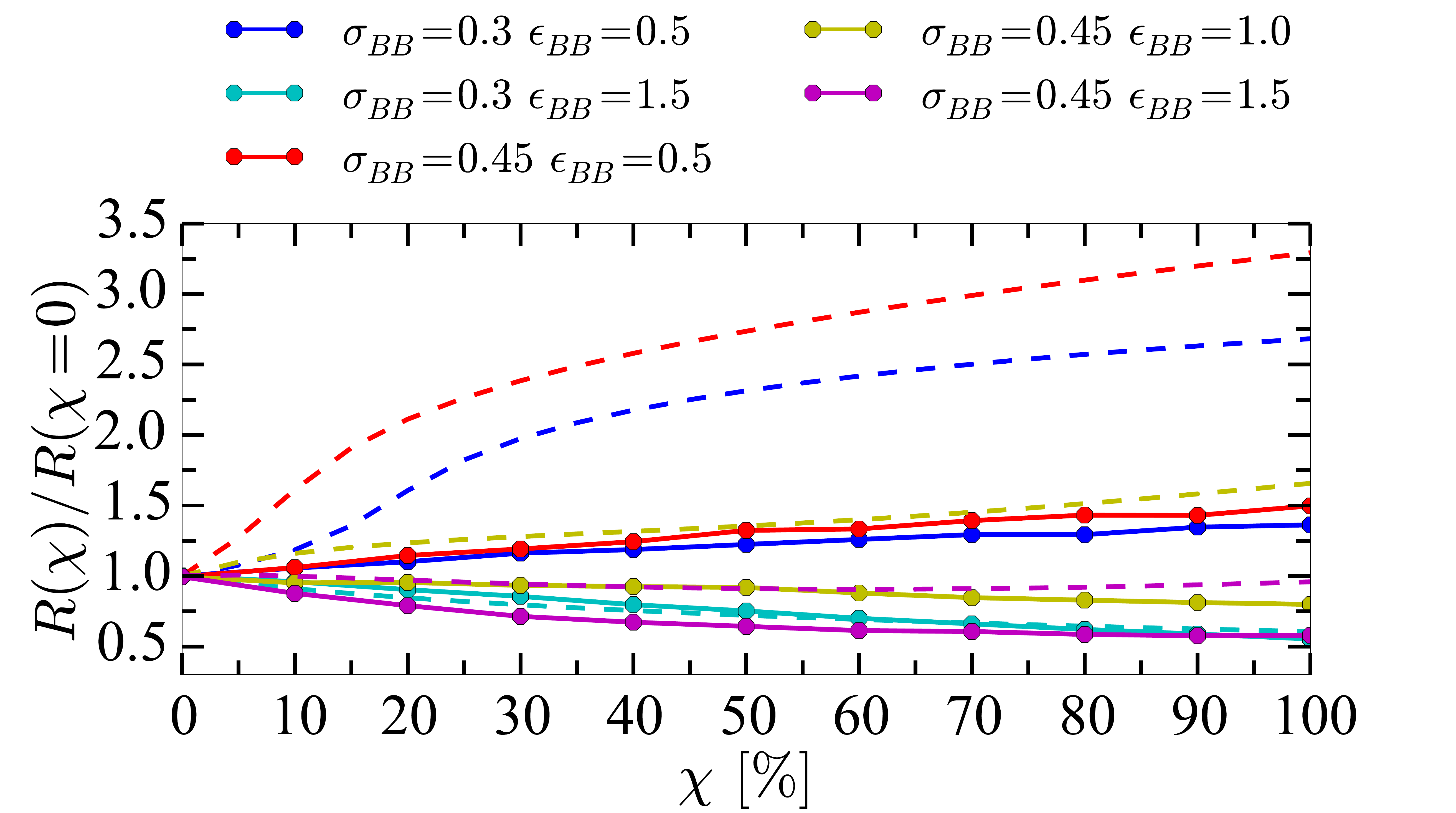}
        \includegraphics[width=8cm]{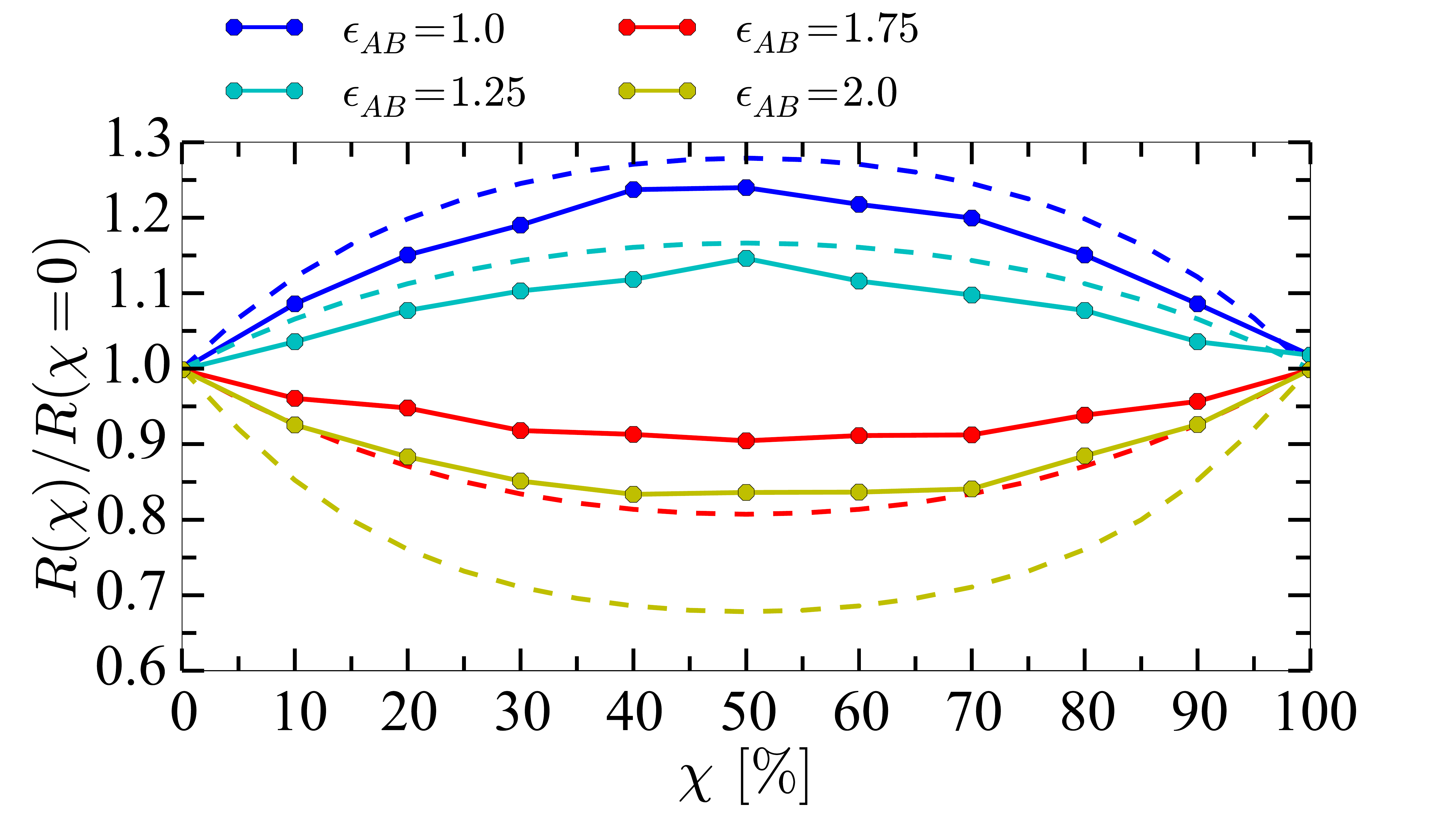}
    \caption{Scaled end-to-end distance $R(\chi)/R(0)$ as a function of copolymerization $\chi$ for different interaction parameters, see legends. 
    Dashed lines are obtained from minimizing the Flory free energy, while symbols depict the simulation results (with solid lines guiding the eye). (top) Additive Lorentz-Berthelot mixing rules apply with $\sigma_{AA}=0.3$ nm, $\epsilon_{AA}=1$ kJ/mol, at $T=300$ K, see Table~I.  (bottom) Non-additive monomer cross-interactions apply 
    with $\sigma_{AA}=\sigma_{BB}=\sigma_{AB}=0.45$~nm at $T=500$~K and $\epsilon_{AA}=\epsilon_{BB}=1.50$~kJ/mol, see Table~II. \label{fig:R_chi}}
\end{figure}

In Fig.~\ref{fig:R_chi} we present theory and simulation data for the polymer size $R(\chi)$ scaled by the respective reference size $R(0)$ as a function of copolymerization $\chi$ and a fixed temperature $T=300$~K.  In the top panel, we plot results for the Lorentz-Berthelot systems summarized in Table~I.  We observe that the polymer size in these additive systems changes monotonically with $\chi$. As expected, copolymerization with monomers B that are more (less) attractive than A lead to 
more collapsed (more swollen) states of the polymer in the simulation. The theory follows these trends, apart from the set $\sigma_{BB} = 0.45$~nm and $\epsilon_{BB}=1.0$~kJ/mol, where the simulation predicts shrinking while theory predicts swelling. In that special case, A and B have the same interaction
depth ($\epsilon_{ij}=1.0$~kJ/mol) but different monomer sizes. Increasing the monomer size has two effects: increasing the excluded volume  repulsion, while at the same time increasing the van der Waals attraction. Interestingly, the balance between those seems subtle enough to not be accurately captured by our mean-field Flory approach that is based only on an approximative virial expansion.  
(Note that for the LJ potential the reduced second virial coefficient $B_2/\sigma^3$ does depend only on the value of $\epsilon$, not that of $\sigma$, cf. also Fig. 2.)

In contrast to the additive models, the non-additive models, shown in the bottom panel of Fig.~\ref{fig:R_chi} for a temperature $T=500$~K, exhibit a non-monotonical behavior for $R(\chi)$. 
Here, we intentionally chose the two limiting cases $\chi=0$ and $\chi=1$ to possess the same interactions (and thus same $T_c)$ to clearly demonstrate the effect.  Since the cross interactions are very different than the A-A and B-B interactions, copolymerization must lead to swelling (shrinking) for more repulsive (more attractive) cross interactions. The strength of the cross-interaction term $\epsilon_{AB}$ is maximal for $\chi=0.5$, and hence 
$R(\chi)/R(0)$ attains an extremum at this value. Both theory and simulations qualitatively agree for all parameter sets. 

\subsection{Critical Temperature\label{sec:Tc_results}}
  
\begin{figure}[b]
    \includegraphics[width=8cm]{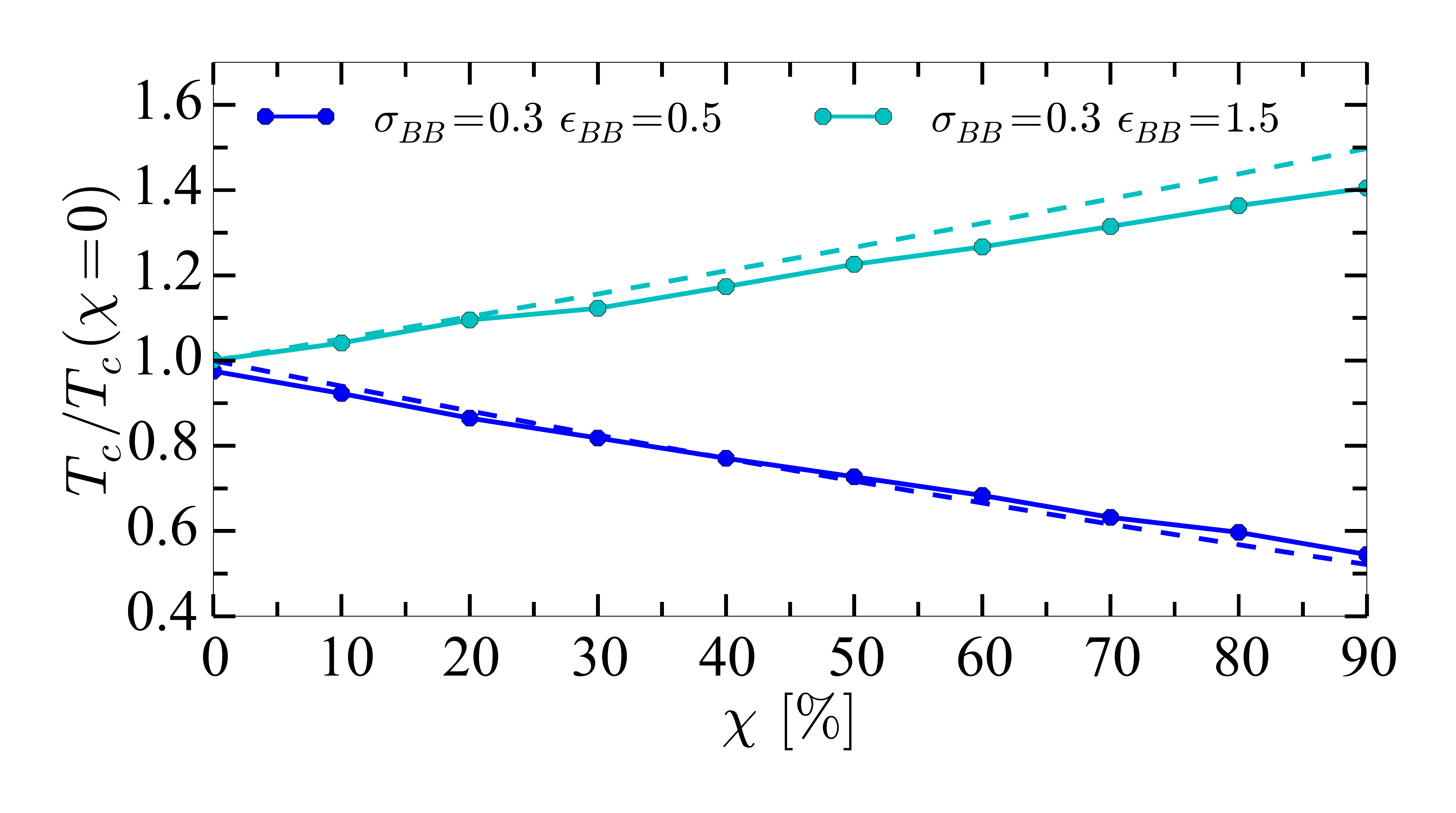}
    \includegraphics[width=8cm]{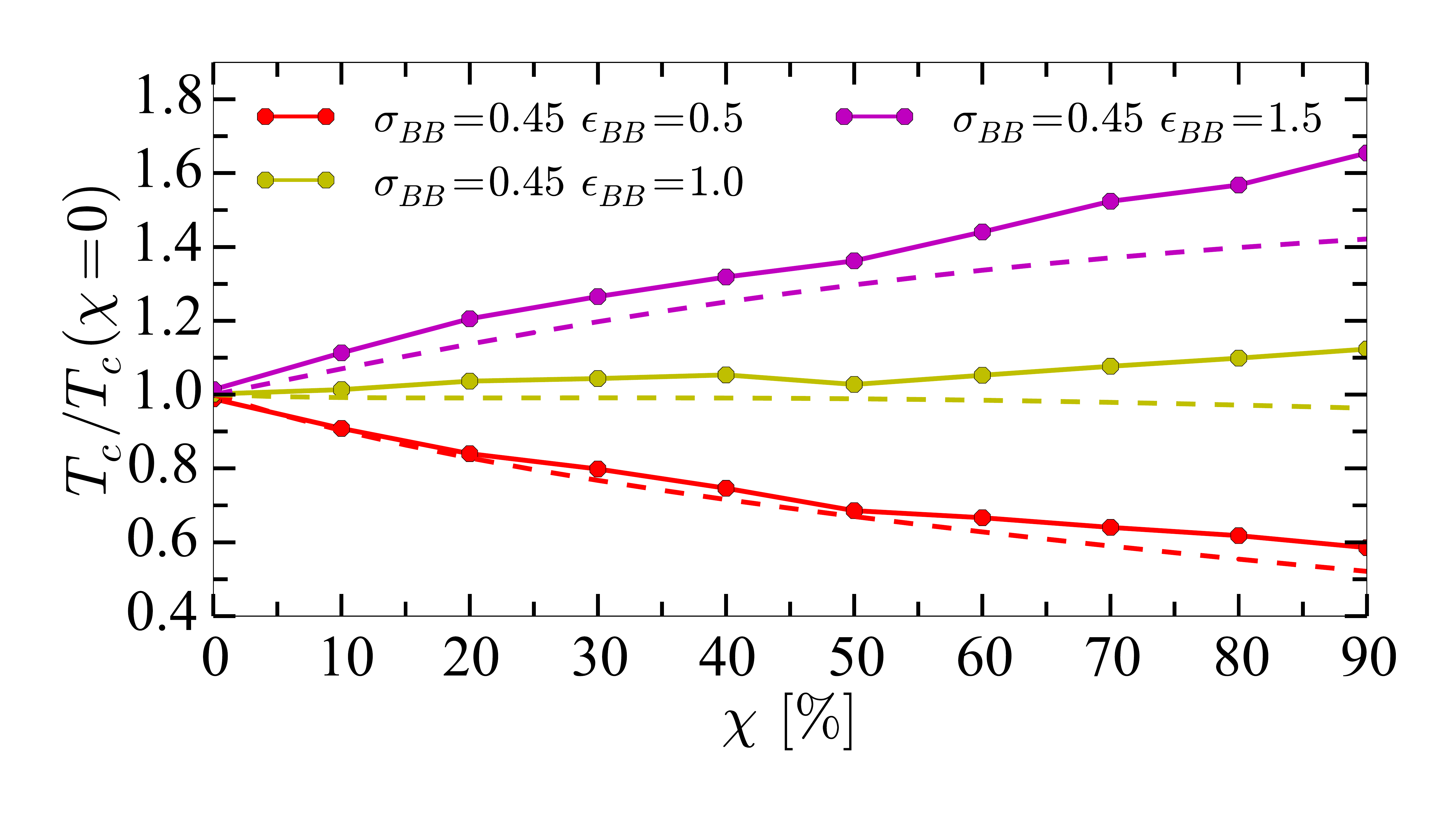}
    \caption{Critical temperature $T_c(\chi)/T_c(0)$ as a function of copolymerization $\chi$ for the additive interaction sets, cf. Table I.  
    The symbols are simulation results with the solid lines guiding the eyes,  while the dashed lines are from Flory theory. 
    Hompolymer A has interactions $\sigma_{AA}=0.3$ nm and $\epsilon_{AA}=1$ kJ/mol and the B-B interactions are according 
    to the legends. \label{fig:Tc_conc}}
\end{figure}

Let us now focus our discussion on the change of the critical temperature $T_c(\chi)$ with copolymerization~$\chi$. In all cases we compare the simulations
directly to the Flory theory and plot the results scaled by the respective reference value $T_c(0)$ (that is, of a homopolymer of type A) 
for a better comparison of the qualitative trends with $\chi$.   In Fig.~\ref{fig:Tc_conc} we present the results for the additive systems.
The results both for theory and simulations show that copolymerizing with more attractive monomers ($|\epsilon_{BB}|>|\epsilon_{AA}|$) increases the critical temperature (UCST) and vice versa for $|\epsilon_{BB}|<|\epsilon_{AA}|$. This is the intuitive behavior since polymers favor the collapsed (swollen) state for highly attractive (repulsive) monomer-monomer interactions, as discussed in the previous section (cf. Fig.~4).  The behavior
can be understood from the expansions eq.~(8) and eq.~(11). Let us focus for now only at the curves with positive slope in the top panel of Fig.~\ref{fig:Tc_conc} where the B-B interaction $\epsilon_{BB} = 1.5$~kJ/mol is larger than the $\epsilon_{AA} = 1.0$~kJ/mol of the A-A interaction. 
The difference in virial coefficients $B_2^{AB}-B_2^{AA}$ is negative for a larger B-B attraction because then also the A-B cross interaction
is more attractive and $B_2^{AB} < B_2^{AA}$. Since the entropy for the transition to the collapsed state is negative, $\Delta S_0<0$, $T_c(\chi)$ increases with 
copolymerization. Analogously, $T_c(\chi)$ decreases for the weaker B-B attraction with  $\epsilon_{BB} = 0.5$~kJ/mol.
The same reasoning holds for all other curves with additive interaction parameters.  

An interesting behavior is again seen for B-B interaction  set $\sigma_{BB} = 0.45$~nm and $\epsilon_{BB}=1.0$~kJ/mol in Fig.~\ref{fig:Tc_conc} (bottom), where only the LJ  size $\sigma_{BB}$ is changed. In this case, the behavior of the polymer size versus $\chi$ in Fig.~4 gave contradicting behavior, because apparently the subtle balance between the increase of both excluded-volume repulsion and van der Waals attraction  with increasing $\sigma_{BB}$ is not well described with our Flory approach.  Analogously, $T_c(\chi)$ shows opposite behavior in theory and simulation, where, consistent with the swelling behavior, the simulation shows a slight increase of $T_c$ and the theory a slight decrease.  However, the  changes in  $T_c(\chi)$ in the full $\chi$ range are relatively small on the shown scale, indicating that the repulsive  and attractive effects almost cancel out in the $T_c(\chi)$ behavior for increasing $\sigma_{BB}$ in this special case. 

\begin{figure}[b]
  \includegraphics[width=8cm]{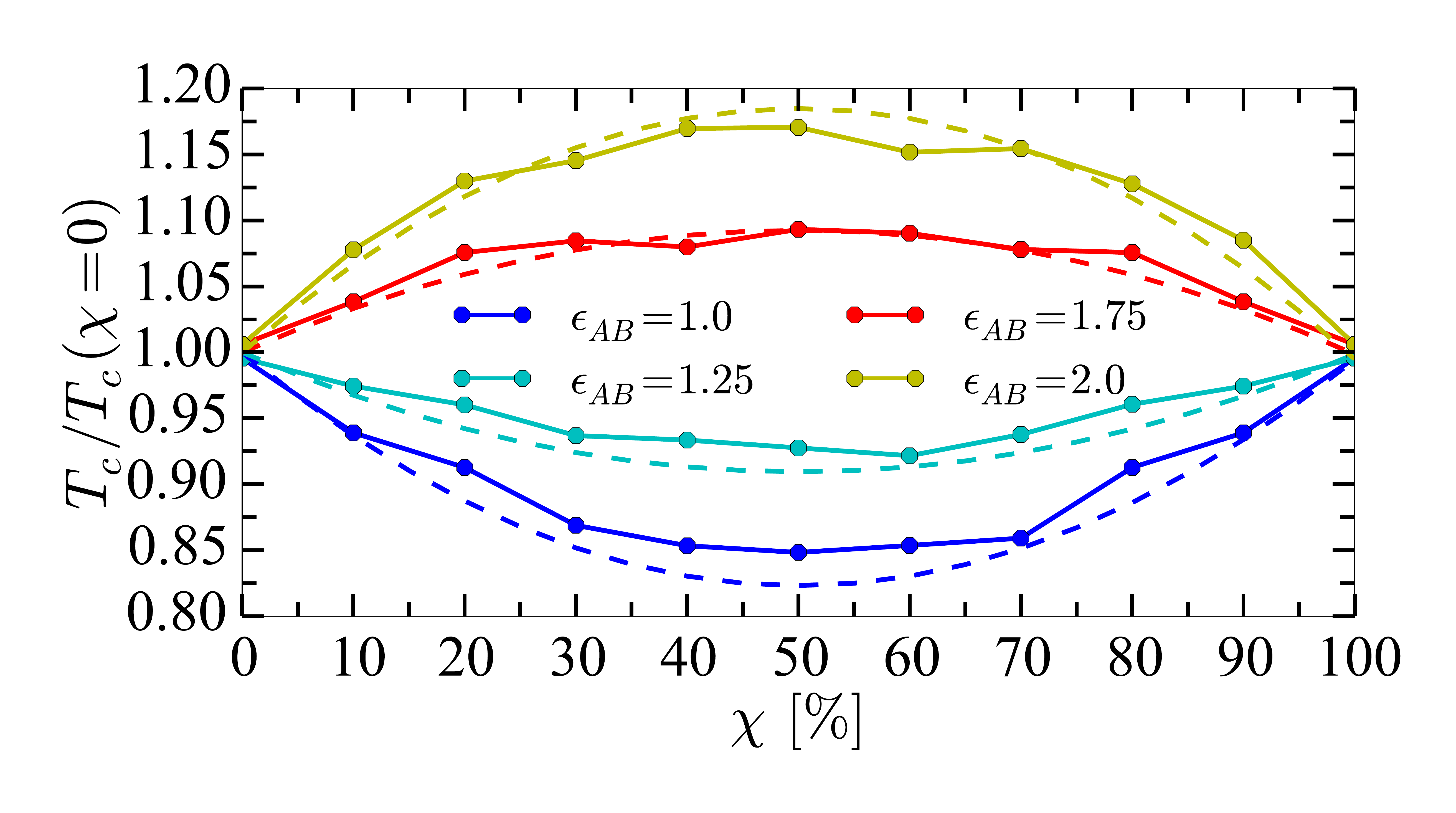}
  \includegraphics[width=8cm]{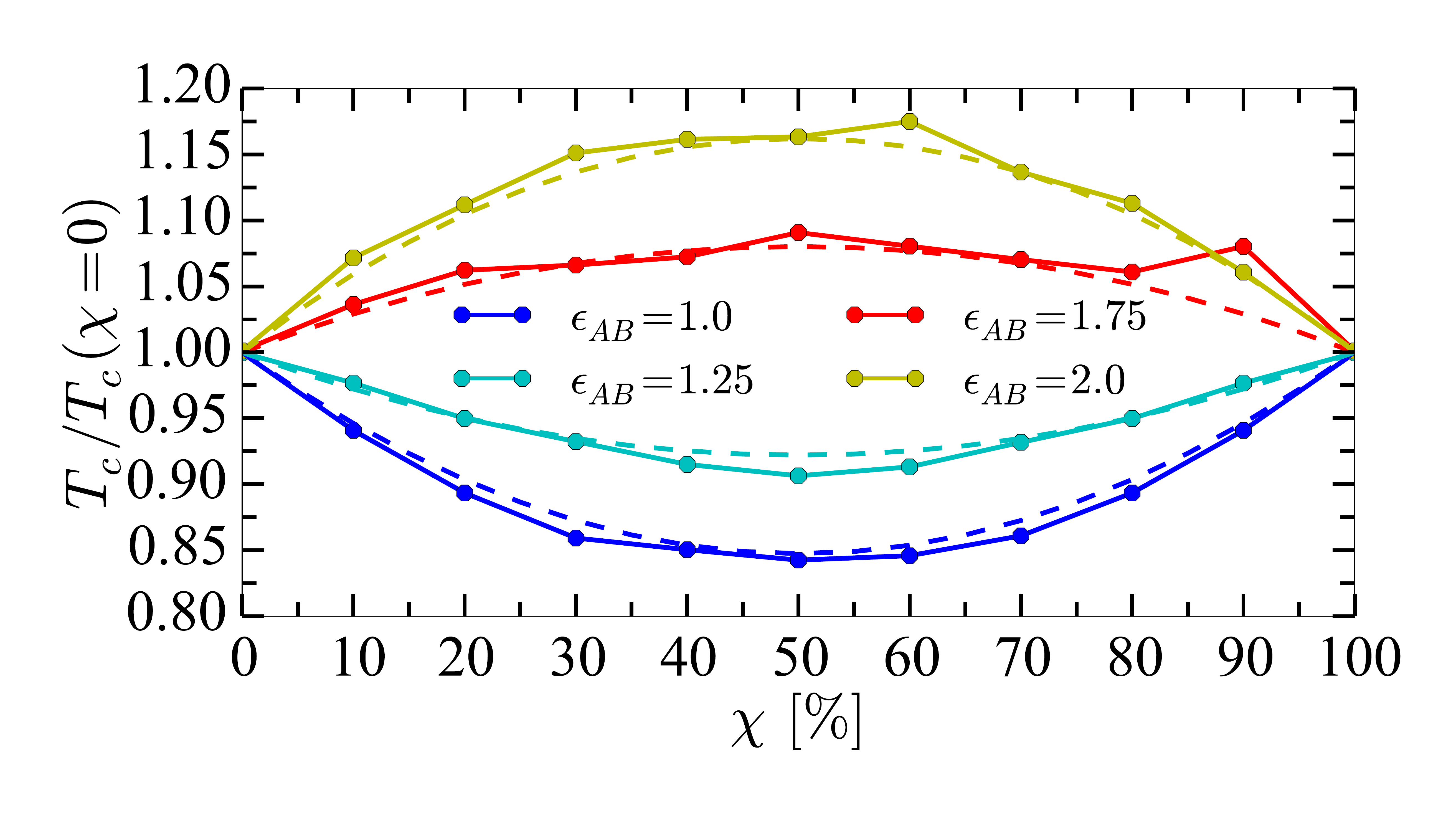}
  \caption{Critical temperature $T_c(\chi)/T_c(0)$ as a function of copolymerization $\chi$ for the non-additive interaction sets, cf. Table II. 
    The symbols are simulation results with the solid lines guiding the eyes,  while the dashed lines are from Flory theory.  
    The cross interactions are according to the legend, while for the equal monomers it is  $\epsilon_{AA}=\epsilon_{BB}=1.5$ kJ/mol and $\sigma_{AA}=\sigma_{BB}=\sigma_{AB}=0.3$~nm, in the top figure and $\sigma_{AA}=\sigma_{BB}=\sigma_{AB}=0.45$~nm, in the bottom figure. \label{fig:Tc_vs_chi_non_Lorentz}}
\end{figure}

In Fig.~\ref{fig:Tc_vs_chi_non_Lorentz} the results for the non-additive systems prepared according to Table \ref{tab:setup_non_Lorentz} are presented. As expected from Fig.~\ref{fig:R_chi},  a nonlinear, strongly non-monotonic dependence of $T_c$ on $\chi$ is observed. This behavior can be understood again by the difference in the virial coeffcients $B_2^{AB}-B_2^{AA}$: since $B_2^{AA} = B_2^{BB}$ now, for a more attractive
$B_2^{AB} < B_2^{AA}= B_2^{BB}$ (and the negative transition entropy), $T_c(\chi)$ must increase symmetrically with the same slope on both sides, for increasing $\chi$ at $\chi=0$, and for decreasing $\chi$ at $\chi=1$. The opposite behavior is seen for a more repulsive $B_2^{AB}>B_2^{AA}= B_2^{BB}$.  The critical temperature $T_c$ is  extremal for $\chi=50$\%\,, where a maximum of $T_c$ is obtained for $\epsilon_{AB}>\epsilon_{AA}=\epsilon_{BB}$ and a minimum for $\epsilon_{AB}<\epsilon_{AA}=\epsilon_{BB}$. The maximum or minimum themselves can be increased or decreased by tuning $\epsilon_{AB}$, respectively.  We note that similar trends can be expected by introducing non-additivity by  $\sigma_{AB}\neq(\sigma_{AA}+\sigma_{BB})/2$ as long as the resulting change in $B_2^{AB}$ versus $B_2^{AA}$ and $B_2^{BB}$ is significant. An example of non-monotonic behavior has been indeed observed  experimentally~\cite{keerl, plamper2012} where preferential hydrogen bonding between NIPAM and N,N-diethylacrylamide (DEAAM) 
monomers is reported.  We note again that for a polymer exhibiting a  LCST all trends are expected to be inverse.

\section{Summary and Concluding Remarks}

In summary,  we have explored how the critical temperature $T_c$ (CST) changes with statistical copolymerization $\chi$ by exploring the size behavior of a single generic polymer model in implicit-solvent computer simulations. A two-component Flory-de Gennes model could describe all trends observed in the simulations, in particular both the monotonic and non-monotonic behavior of $T_c(\chi)$, that we have demonstrated to explicitly reflect the degree of non-additivity of the monomeric cross interactions. The discussed trends resemble the miscibility behavior in copolymer blends~\cite{barlow} and are consistent with experimental LCST behavior of thermosensitive polymers. For linear and uncharged copolymers experiments have reported both, a monotonic \cite{keerl, plamper2012,maeda2009,park2007,glassner2014,djokpe2001} as well as a non-monotonic \cite{keerl, plamper2012,maeda2009,park2007} dependency of $T_c$ on the degree of copolymerization $\chi$.  We demonstrated that the non-monotonicity can be explained by the non-additive contributions from the preferential  attraction between NIPAM and N,N-diethylacrylamide (DEAAM) monomers, where unlike acceptor and donor pairs attract by H-bonding mechanisms. Our effective mean-field treatment is thus complementary to the more microscopic explanation of these effects put forward recently on the basis of cooperative hydration effects.~\cite{tanaka2013} 

Obviously, there are many limitations of our work that could be addressed in future. In the presented work, for instance, a temperature-independent Lennard-Jones potential has been used. However, more realistic effective potentials should include solvent and maybe even cosolvent effects. To directly connect to LCST experiments, it would be necessary to introduce explicitly temperature-dependent pair potentials~\cite{schellman, paschek, makowski} and virial coefficients.~\cite{reinhardt2013fine} Other effects  not considered in our model are for example the sequential arrangement of the co-monomers, i.e., the composition effect~\cite{balazs} or electrostatic charging effects.~\cite{kawasaki}  For the latter, a simple Donnan-like model was devised by us recently where experimental LCST changes by charge fractionation of the polymer and the addition of salt could be described in a satisfactory fashion.~\cite{heyda2014a}
\vspace*{11pt}

\section{Acknowledgments}

The authors are grateful to Felix Plamper for inspiring discussions. 
J.H. and J.D. acknowledge funding from the Alexander-von-Humboldt (AvH) Stiftung, Germany.
R.C. and J.D. acknowledge funding from the Deutsche Forschungsgemeinschaft (DFG), Germany.  
J.D. is very thankful for  funding from  the ERC (European Research Council) within the Consolidator Grant with project number 646659 - NANOREACTOR.

%


\end{document}